\documentstyle[preprint,aps]{revtex}  

\begin{document}

\title{Spin softening in models with competing interactions:\\
a new high anisotropy expansion to all orders}
\author{ F. Seno and J.M. Yeomans}
\address{Department of Physics, University of Oxford, 1
Keble Road, Oxford, OX1 3NP, U.K.}
\maketitle
\begin{abstract}
An expansion in inverse spin anisotropy, which enables us to study the
behaviour of discrete spin models as the spins soften, is developed.
In particular we focus on models, such as the chiral clock model and
the $p$-state clock model with competing first and second neighbour
interactions, where there are special multiphase points at zero
temperature at which an infinite number of ground states are degenerate.
The expansion allows calculation of the ground state phase diagram
near these points as the spin anisotropy, which constrains the spin to
take discrete values, is reduced from infinity. Several different
behaviours are found, from a single first order phase boundary to
infinite series of commensurate phases.
\vspace{1. truecm}
\bgroup\draft\egroup
\end{abstract}

Competing interactions lead to many important physical phenomena
\cite{Selke92}.
Examples are charge density waves, long-period phases in binary alloys
and ferrimagnetism in the rare earths \cite{Loiseau85,Jensen91}. Considerable
understanding of
these systems has been obtained by studying one-dimensional models
which embody the competition. The most famous of these is the
Frenkel-Kontorova model, atoms connected by springs lying in a
periodic potential\cite{Frenkel}. Several spin systems behave similarly, for
example
the chiral X-Y model with $p$-fold spin anisotropy,
$D$\cite{Taylor84,Chou86,Yokoi88,GrifR}. Here the
value of the chirality selects a given wavevector which competes with
the preferred spin directions defined by $D$ .

These models have subtle and complicated ground states. Long-period
commensurate and incommensurate phases are important and devil's
staircase behaviour, upsilon points and symmetry-breaking second order
transitions are among the features observed \cite{Aubry78,Bas82,Sas92}. If the
interactions are
convex the ground state behaviour is rather well
understood\cite{Aubry83a,Aubry83b}. For
non-convex interactions, however, much less is known and most work has
been numerical\cite{Chou86}. Spin systems, which generically fall into this
category, have received less attention than Frenkel-Kontorova
models\cite{Sas92,Seno94}.
An important aim is to understand which ground state features can
occur in these models and to ascertain whether any universal behaviour
arises.

To this end we have developed a new analytic approach which allows us to
study the behaviour of spin models in the limit where the pinning
potential which results from the spin anisotropy, $D$, is large. This is an
expansion in inverse spin anisotropy which can be carried to all
orders where necessary. In the limit of infinite $D$ the
spins can take only discrete values and the ground state typically
consists of a few short-period phases as a controlling parameter such
as the chirality is varied.  Interesting behaviour can occur
when the boundary between different ground states is infinitely
degenerate, at so-called multiphase points \cite{FS80,Aubry2}. Here, as $D$
decreases from infinity and the spins are allowed to soften, we are
able to demonstrate several different behaviours, ranging from a
single first order boundary to infinite series of commensurate phases.

The approach is described for the one-dimensional, classical X-Y model with
competing first- and second-neighbour interactions and $p$-fold spin
anisotropy

\begin{equation}
{\cal H}=\sum_{i}\bigg\{ -J_{1}\cos (\theta_{i-1}-\theta_i)
+J_{2}\cos (\theta_{i-2}-\theta_i)
-{D}(\cos p\theta_i-1)/p^2 \bigg\}.
\label{hamil}
\end{equation}
$J_{1}$, $J_{2}$ and $D$ are chosen to be positive and $\theta_{i}$ is
an angle between $0$ and $2\pi$.
For $D=0$ the ground state of the Hamiltonian\ (\ref{hamil})
is ferromagnetic for $x \equiv J_2/J_1 \le 1/4$ and
modulated with a wavevector $2 \pi q=\cos^{-1} (1/4x)$ for $x>1/4$.
At $D=\infty$ (\ref{hamil})
becomes a $p$-state clock model and $\theta_i$ can take values $2 \pi
n_i/p$, $n_i=0,1,2 \ldots p-1$. The ground state is now a
sequence of short-period phases as $x$ is varied.
The boundaries between the different ground states can either be
simple points where only a few distinct phases have the same energy
or multiphase points where an infinite number of different states are
degenerate.

A particular case of the latter which has very rich behaviour is the boundary
between
states with wavevectors $1/p$ $(\{n_i\}= \ldots012345\ldots)$ and $3/(2p)$
$(\ldots013467\ldots)$ which occurs for all
$p\ge6$ at
\begin{equation}
x_0=\{\cos (2 \pi /p)-\cos (4 \pi /p)\}/
\{2(\cos (4 \pi /p)-\cos (6 \pi /p))\}.
\end{equation}
Here all phases with $\delta n_i \equiv n_{i+1}-n_{i}=1,2$, with the
proviso that
$\delta n_i =\delta n_{i+1}=2$ is forbidden are degenerate for $D=0$.

To facilitate
description of the ground-state structures near this point it is
helpful to define a wall as a position where $\delta n_i=2$ and a band
as a sequence of spins between walls \cite{FS80,SRY93}.
 Then we label a state which is
made up of a repeating sequence of bands $m_1,m_2 \ldots m_n$ as
$\langle m_1,m_2 \ldots m_n \rangle$. With this definition the phases
bordering the multiphase point at $x_{0}$ are $\langle \infty \rangle$
and  $\langle 2 \rangle$ and all phases consisting of bands of length
$\ge 2$ are stable at the point itself.

The aim here is to investigate the stability of the phase diagram
around $x_{0}$ as $D$ decreases from infinity as an expansion in
$D^{-1}$. Although we focus on the Hamiltonian (\ref{hamil}) and a
particular series of multiphase points the method is general and
results for other models will be described later in the paper.

The expansion is possible because for $D$ large the spins deviate from
their
clock positions by an angle analytic in $D^{-1}$. Writing
\begin{equation}
\theta_i=\theta_i^0+\tilde{\theta}_i
\end{equation}
and keeping only terms quadratic in $\tilde{\theta}_{i-1}-\tilde{\theta}_{i}$
and $\tilde{\theta}_{i-2}-\tilde{\theta}_{i}$ in the expansion of the
Hamiltonian (\ref{hamil})
gives
\begin{eqnarray}
\tilde{\cal H}= &&{\cal H}\bigg|_{ D=\infty}+J_1 c_{i,1}
(\tilde{\theta}_{i-1}-\tilde{\theta_i}+{s_{i,1}}/{c_{i,1}})^2 \big / 2
- J_1 {s_{i,1}^2}/ 2 {c_{i,1}}\nonumber \\
&&-J_2 c_{i,2} (\tilde{\theta}_{i-2}-\tilde{\theta_i}+{s_{i,2}}/{c_{i,2}})^2
\big / 2 + J_2 {s_{i,2}^2}/ 2  {c_{i,2}}
+D\tilde{\theta_i}^2/2
\label{quadham}
\end{eqnarray}
where
\begin{eqnarray}
s_{i,1}=\sin{(\theta_{i-1}^{0}-\theta_{i}^{0})},&& \ \ \ \
c_{i,1}=\cos{(\theta_{i-1}^{0}-\theta_{i}^{0})},\nonumber\\
s_{i,2}=\sin{(\theta_{i-2}^{0}-\theta_{i}^{0})},&& \ \ \ \
c_{i,2}=\cos{(\theta_{i-2}^{0}-\theta_{i}^{0})}.
\label{screfs}
\end{eqnarray}

The equilibrium values of the $\tilde{\theta}_{i}$ are given by
minimising the Hamiltonian (\ref{quadham}).
This leads to linear recursion relations
\begin{eqnarray}
\tilde{\theta}_{i}=&&J_1 \{c_{i,1}(\tilde{\theta}_{i-1}-\tilde{\theta}_i)

+c_{i+1,1}(\tilde{\theta}_{i+1}-\tilde{\theta}_i)
                                +s_{i,1}-s_{i+1,1}\}/D\nonumber \\
&&-J_2 \{c_{i,2}(\tilde{\theta}_{i-2}-\tilde{\theta}_i)

+c_{i+2,2}(\tilde{\theta}_{i+2}-\tilde{\theta}_i)
                                +s_{i,2}-s_{i+2,2}\}/D.
\label{rr}
\end{eqnarray}
If the full Hamiltonian (\ref{hamil}) is used non-linearities
 appear in the recursion relations (\ref{rr}).
However, these do not affect the
leading order  terms needed   for the subsequent calculations.

Writing
\begin{equation}
\tilde{\theta}_i=\frac{\tilde{\theta}_{i}^{1}}{D}+\frac{\tilde{\theta}_{i}^{2}}{D^2}+\ldots
+\frac{\tilde{\theta}_{i}^{n}}{D^n}+\ldots
\end{equation}
equation (\ref{rr}) immediately gives
\begin{eqnarray}
\tilde{\theta}_{i}^{1}&=&J_1
(s_{i,1}-s_{i+1,1})-J_2(s_{i,2}-s_{i+2,2})\label{initial},\\
\tilde{\theta}^{n}_{i}&=&J_1\{c_{i,1}(\tilde{\theta}_{i-1}^{n-1}-\tilde{\theta}_i^{n-1})
 +c_{i+1,1}(\tilde{\theta}_{i+1}^{n-1}-\tilde{\theta}_i^{n-1})\}\nonumber\\
&&-J_2 \{c_{i,2}(\tilde{\theta}_{i-2}^{n-1}-\tilde{\theta}_i^{n-1})

+c_{i+2,2}(\tilde{\theta}_{i+2}^{n-1}-\tilde{\theta}_i^{n-1})\}.
\label{ho}
\end{eqnarray}
It follows from equations (\ref{screfs}) and (\ref{initial}) that the two
spins in a 2-band, the edge spins of a 3-band and the two spins nearest
each edge
 of a band of length $\ge 4$ have a deviation
${\cal O}(1/D)$.
All other spins remain unmoved to this order.
{}From equation (\ref{ho}) it is apparent that the next pairs of spins moving
in from each edge of the band will have a deviation ${\cal O}(1/D^2)$; the
next pairs ${\cal O}(1/D^3)$ and so forth.

We establish the stable phase sequences near $x_{0}$ by following an inductive
argument originally due to Fisher and Selke \cite{FS80} (see also
\cite{Villain,Aubry1}). Defining $E_{\langle
\alpha \rangle}$ as the ground state energy per spin of $\langle
\alpha \rangle$ and $n_{\langle \alpha \rangle}$ as the number of spins
per period, this can be summarised as
follows: assume that ${\cal O} (1/D^{n})$ two neighbouring phases $\langle
\alpha \rangle$ and
$\langle \beta \rangle $ are stable and all phases comprised
of $\alpha$- and $\beta$-sequences are degenerate on the boundary
between them. Then the first phase that can appear between them
is $\langle \gamma \rangle \equiv \langle \alpha \beta\rangle$.
If $\Delta E \equiv n_{\langle \alpha \beta\rangle} E_{\langle \alpha
\beta\rangle}
-n_{\langle \alpha \rangle} E_{\langle \alpha \rangle}
-n_{\langle \beta\rangle} E_{\langle \beta\rangle}
> 0$ the boundary remains stable to all
orders. If $\Delta E < 0$ and  ${\cal O}(1/D^m)$ with $ m > n$, however,
$\langle \alpha \beta\rangle $ appears as a stable phase on the
$\langle \alpha \rangle:\langle \beta \rangle$ boundary over a region
${\cal O} (1/D^{m})$ and the analysis must recommence about the new $\langle
\alpha \rangle : \langle \alpha \beta \rangle$ and
$\langle \alpha \beta \rangle: \langle \beta \rangle$ boundaries.

Hence the task is to calculate $\Delta E$. Let
$n_{\langle \alpha \rangle}=n_1$, $n_{\langle \gamma \rangle}=n$ and label
the repeating spin
sequences of $\langle \alpha \rangle$, $\langle \beta \rangle$ and
$\langle \gamma \rangle$ as $(\alpha_1,\alpha_2\ldots\alpha_{n_{1}})$,
$(\beta_{n_{1}+1},\beta_{n_{1}+2}\ldots\beta_{n})$ and
$(\gamma_1,\gamma_2\ldots\gamma_n)$ respectively.
It is lengthy but not difficult to show that
\begin{eqnarray}
\Delta E&=&
 J_1 c_{1,1} \{(\alpha_0-\beta_0)(\gamma_1-\gamma_{n_{1}+1})
 -(\alpha_1-\beta_1)(\gamma_0-\gamma_{n_{1}})\}/2 \nonumber\\&&
- J_2 c_{1,2} \{(\alpha_{-1}-\beta_{-1})(\gamma_1-\gamma_{n_{1}+1})
-(\alpha_1-\beta_1)(\gamma_{-1}-\gamma_{n_{1}-1})\}/2 \nonumber\\&&
- J_2 c_{2,2} \{(\alpha_0-\beta_0)(\gamma_2-\gamma_{n_{1}+2})
-(\alpha_2-\beta_2)(\gamma_0-\gamma_{n_{1}})\}/2
\label{energydiff}
\end{eqnarray}
This formula is exact for the quadratic Hamiltonian~(\ref{quadham}).
Higher order terms in the full Hamiltonian~(\ref{hamil}) appear as
higher order corrections.

The value of  $\Delta E$ must obviously be  independent of the choice of
spin labelling.
However, given an appropriate choice of labelling  only the leading order terms
in the spin differences need be calculated.
 This follows from noting that all possible states have an axis of
symmetry. This lies either on or between spins depending on whether the number
of spins in a period is even or odd.
 When states are
combined there are two possibilities:
(i) $n_{\langle \alpha \rangle}$ odd, $n_{\langle \beta \rangle}$ odd
$\rightarrow$ $n_{\langle \gamma \rangle}$ even.
For an odd state symmetry demands that one spin remains fixed
($\tilde{\theta}=0$).
Therefore we may choose $\alpha_0 =0$, $ \beta_{0}=0$.
(ii) $n_{\langle \alpha \rangle}$ odd, $n_{\langle \beta \rangle}$ even
$\rightarrow$
$n_{\langle \gamma \rangle}$ odd.
We choose $\alpha_{(n_{1}+1)/2}=0$ or equivalently
$\gamma_{1}-\gamma_{n_{1}+1}=
\gamma_{0}-\gamma_{n_{1}}$. This implies $(\alpha_1 -\beta_1
)=-(\alpha_{0}-\beta_{0})$.
(Consideration of how neighbouring states are constructed shows that
two adjacent even states are never generated.)

The spin differences can be calculated to leading order by replacing
$\theta_i$ with $(\alpha_i -\beta_i )$ or the closely related
$(\gamma_i -\gamma_{n_{1}+i})$ in
equations (\ref{initial}) and (\ref{ho}). Let $(\alpha_i
-\beta_i)^{1}=0$,  $i<n_0$. The choice of spin labelling detailed
above maximises $n_{0}$. Iteration of the recursion
equations leads after an involved calculation which will be detailed elsewhere
\cite{SY94} to the following
expressions for the energy differences for $n>0$

\noindent ${{{(i)\; \underline{ n_{\langle \gamma \rangle}=1\;[{\rm mod}\; 4]
\equiv 4n+1;\;\;\;\;n_{0}=2n-1}}}}$
\begin{eqnarray}
\Delta
E&=&\{-J_{1,1}J_{2}^{2n}c_{3}^{2\tilde{n}_w}c_{2}^{2n-2-2\tilde{n}_w}(s_3-s_2)^2\nonumber\\
&&-2J_{1,2}J_{2}^{2n-1}J_{1}c_{3}^{n_{w}+\tilde{n}_w-1}c_{2}^{2n-3-n_{w}-\tilde{n}_w}
(s_3-s_2)^2(c_3c_1(n-1-x)+c_{2}^{2}x)\nonumber\\
&&-2 J_{1,2}J_{2}^{2n-1}J_1
c_{3}^{n_{w}+\tilde{n}_w}c_{2}^{2n-2-n_{w}-\tilde{n}_w}
(s_3-s_2)(s_2-s_1)\}/D^{2n}
\label{start1}
\end{eqnarray}

\noindent ${(ii)\; \underline{n_{\langle \gamma \rangle}=2\;[{\rm mod}\; 4]
\equiv 4n+2;\;\;\;\;n_{0}=2n-1}}$
\begin{eqnarray}
\hspace{-4 truecm}  \Delta E&=&
\{J_{2}^{2n+1}c_{3}^{2\tilde{n}_w}c_{2}^{2n-1-2\tilde{n_w}}(s_3-s_2)^2\}/D^{2n}
\label{start2}
\end{eqnarray}

\noindent ${(iii)\; \underline{n_{\langle \gamma \rangle}=3\;[{\rm mod} \; 4]
\equiv 4n+3;\;\;\;\;n_{0}=2n}}$
\begin{eqnarray}
\hspace{-0.5 truecm} \Delta E&=&
\{2J_{1,2}J_{2}^{2n}J_{1}c_{3}^{n_{w}+\tilde{n}_w-1}c_{2}^{2n-2-n_{w}-\tilde{n}_w}
(s_3-s_2)^2(c_3c_1(n-\tilde{x})+c_{2}^{2}\tilde{x})\nonumber\\
&&+2J_{1,2}J_{2}^{2n}J_1
c_{3}^{n_{w}+\tilde{n}_w}c_{2}^{2n-1-n_{w}-\tilde{n}_w}
(s_3-s_2)(s_2-s_1)\}/D^{2n+1}
\label{start3}
\end{eqnarray}

\noindent ${(iv)\; \underline{n_{\langle \gamma \rangle}=4\; [{\rm mod }\; 4]
\equiv 4n+4;\;\;\;\;n_{0}=2n}}$
\begin{eqnarray}
\hspace{-3.8 truecm} \Delta E&=&
\{-J_{2}^{2n+2}c_{3}^{2\tilde{n}_w}c_{2}^{2n-2\tilde{n}_w}
(s_3-s_2)^2\}/D^{2n+1}
\label{start4}
\end{eqnarray}
where $c_{m}=\cos(2 \pi m/p)$, $s_{m}=\sin (2 \pi m/p)$,
$J_{1,1}=J_{1} \cos (2 \pi (\alpha_1^{0} -\alpha_0^{0} )/p)$,
$J_{1,2}=J_{2} \cos (2 \pi (\alpha_1^{0} -\alpha_{-1}^{0} )/p)$,
$J_{2,2}=J_{2} \cos (2 \pi (\alpha_2^{0} -\alpha_{0}^{0} )/p)$,
$n_{w}$ and $\tilde{n}_{w}$ are the number of walls between $n_{0}$ and
$2$ and  $n_{0}$ and $1$  respectively, $x=\sum_{i=2,4\ldots(2n-2)}(\delta
n_i-1)$ and $\tilde{x}=\sum_{i=1,3\ldots(2n-1)}(\delta n_i-1)$.

Different formulae are needed for  for the phases
$\langle m \rangle$ which border $\langle \infty \rangle$

\noindent${(i)\;  \underline{\langle 4n \rangle + \langle \infty \rangle
\rightarrow \langle
4n+1 \rangle}}$
\begin{eqnarray}
\Delta E& = &
-J_{2}^{2n-2}c_{2}^{2n-2}\{p_{1}^{2}((2n-1)J_{1}c_{1}+J_{2}c_{2})
-2p_{1}p_{2}J_{2}c_{2}\}/D^{2n}
\label{end1}
\end{eqnarray}
\noindent${(ii)\; \underline{\langle 4n + 1 \rangle + \langle \infty \rangle
\rightarrow \langle 4n+2 \rangle}}$
\begin{eqnarray}
\hspace{-6.5 truecm} \Delta E& = & p_{1}^2J_{2}^{2n-1}c_{2}^{2n-1}/D^{2n}
\label{end2}
\end{eqnarray}
\noindent${(iii)\; \underline{\langle 4n+2 \rangle + \langle \infty \rangle
\rightarrow \langle
4n+3 \rangle}}$
\begin{eqnarray}
\Delta E& = &
J_{2}^{2n-1}c_{2}^{2n-1}\{p_{1}^{2}(2J_{1}nc_{1}+J_{2}c_{2})
-2J_2p_{1}p_{2}c_{2}\}/D^{2n+1}
\label{end3}
\end{eqnarray}
\noindent${(iv)\; \underline{\langle 4n + 3 \rangle + \langle \infty \rangle
\rightarrow \langle 4n+4 \rangle}}$
\begin{eqnarray}
\hspace{-5.5 truecm} \Delta E& = - & p_{1}^2J_{2}^{2n}c_{2}^{2n}/D^{2n+1}
\label{end4}
\end{eqnarray}
where $p_1 =-J_2 (s_3 -s_2 )$ and $p_{2}=J_1 (s_2 -s_1)-J_2 (s_3 -s_2 )$ .

Results for low order phases, $n=0$, can be obtained directly from
equation~(\ref{quadham})  can be obtained analytically and then the
formulae (\ref{start1})--(\ref{end4}) used to build up the phase diagram
inductively. Although the energy differences $\Delta E$ are cumbersome it is
not hard to establish
their sign for different phase sequences and values of $p$.
If $\Delta E$ is not too small the results can be checked numerically.
This is done by minimizing the ground state energy (\ref{hamil}) with respect
to the $\theta_i$ giving a set of coupled non-linear equations which can be
solved
by iteration. Typically it is feasible to identifiy phases which apper
${\cal O}(1/D^{7})$.

The
results  for the Hamiltonian (\ref{hamil}) are strongly $p$-dependent.\\
$p=6$: the $\langle 2 \rangle : \langle 3 \rangle$ boundary is stable.
All energy differences are negative for phases which can be constructed by the
iterative process of combining neighbouring states which
contain  bands
of length $\ge 3$. Hence all these phases spring from the multiphase
point.\\
$p=7$:  all phases lying between  $\langle 223 \rangle$ and
$\langle \infty \rangle$  are stable. The $\langle 2 \rangle : \langle 223
\rangle$ boundary is not split.\\
$p=8$: many of the energy differences are zero.  Hence the
formalism breaks down. Numerical results show however that at least all
phases expected to appear ${\cal O}(1/D^{5})$ are stable.\\
$p=9$: no  clear pattern emerges. ${\cal O}(1/D^5)$ the phase
sequence is $\langle \infty \rangle : \langle 4 \rangle : \langle 34 \rangle :
\langle 3 \rangle ; \langle 2333 \rangle; \langle 233 \rangle ;
\langle 23 \rangle ; \langle 23223 \rangle; \langle 223 \rangle ;
\langle 2223 \rangle ; \langle 2 \rangle$, where $:$ denotes a stable boundary
and $;$ a boundary which may be split at higher orders of the expansion.\\
$p=10$: many of the energy differences are zero. Numerically we have been
able to show that  ${\cal O}(1/D^{7})$ only the phases
$\langle 2^k3 \rangle, k=1,2 \ldots,5 $  are stable between $\langle 2 \rangle$
and $\langle 3 \rangle$. \\
$p \ge 11$: the $\langle 2 \rangle : \langle \infty \rangle$ boundary
is stable and no new phases appear near $x_{0}$ and the transition is
first order.

Results have also been obtained for several other models. The
chiral X-Y model with $p$-fold anisotropy
\begin{equation}
{\cal H}=\sum_{i} \{ -J \cos (\theta_{i-1}-\theta_i+\Delta)
-D(\cos p\theta_i-1)/p^{2}\}
\end{equation}
becomes the chiral clock model in the $D \rightarrow \infty$ limit.
At the multiphase point at $\Delta=\pi / p$ between the ferromagnetic
$(\ldots 000 \ldots)$ and chiral $(\ldots 012 \ldots )$ states
\begin{equation}
\Delta E=-\{4 \sin^{2}(\pi/p) ( J \cos (\pi/p))^{n_{\langle \gamma
\rangle}}\}
/\{D^{n_{\langle \gamma \rangle}-1}\}
\end{equation}
for a final phase $\langle \gamma \rangle$. This is always negative
for $p \ge 3$ indicating, in agreement with Chou and Griffiths \cite{Chou86}
that all phases
are stable.

The Frenkel-Kontorova model with a piecewise parabolic potential
\begin{equation}
{\cal H}=\sum_{i} \{-J  (\theta_{i-1}-\theta_i+\Delta)^2
-D(\theta_i-\theta_{i}^{0})^2 \}
\end{equation}
 has a multiphase point at $\Delta=\pi/p$
between the ferromagnetic and chiral states. Here
\begin{equation}
\Delta E=-(8 \pi^2 J^{n_{\langle \gamma \rangle}})/(p^2 D^{n_{\langle
\gamma \rangle}-1})
\end{equation}
in agreement with the exact results of Aubry \cite{Aubrynew}.

The series expansion in inverse anisotropy outlined here provides a
new tool to understand the crossover between discrete and continuous
spin models particularly near $D=\infty$ where the narrowness of the
phases renders numerical work difficult. Many interesting avenues
remain to be explored. In  particular it would be of interest to understand
the behaviour of models where the
leading term in the energy differences vanishes and the recursion
relations become non-linear. Moreover it may be  possible to recast the
formalism in
terms of interactions between domain walls \cite{Bas82,Villain,Spzilka} and
hence attempt to
classify the high-$D$ behaviour of systems with modulated structures.
Finally the multiphase point considered here is a special case of the
anti-integrable limits described by Aubry \cite{Aubry2} which also
exist in electronic models and systems of coupled anharmonic
oscillators. Investigation of whether similar expansions exist for
these models would be of considerable interest.

\vspace{1cm}

F.S. acknowledges the European Community for a postdoctoral fellowship
under the Human Capital and Mobility Programme.

#

\begin{references}
\bibitem{Selke92} W. Selke in {\em Phase Transitions and critical phenomena}
vol. 15, eds. C. Domb and J.L. Lebowitz, (New York : Academic , 1992).

\bibitem{Loiseau85} A. Loiseau, G. van Tendeloo, R. Portier and F. Ducastelle,
J. Physique {\bf 46}, 595 (1985).

\bibitem{Jensen91} J. Jensen and A.R. Mackintosh,  {\em Rare Earth Magnetism},
Oxford Univerisity Press, Oxford (1991).

\bibitem{Frenkel} T. Kontorova and Y.I. Frenkel, Phys. Z Sowjetunion {\bf 13} 1
(1938).

\bibitem{Taylor84} A. Banerjea and P.L. Taylor, Phys. Rev. {\bf B30}, 6489
(1993).

\bibitem{Chou86} W. Chou and R.B. Griffiths, Phys. Rev. {\bf B34}, 6219 (1986)

\bibitem{Yokoi88} C.S.O. Yokoi, L.H. Tang and W.R. Chou, Phys. Rev.
{\bf B37}, 2173 (1988).

\bibitem{GrifR} R.B. Griffiths in {\em Fundamental Problems in Statistical
Mechanics VII}, ed. H. van Beijeren (North Holland, Amsterdam , 1990)

\bibitem{Aubry78} S. Aubry, in {\em Solitons and Condensed Matter Physics},
eds. A.R. Bishop and T. Schneider, (Berlin: Springer 1978).

\bibitem{Bas82} K.E. Bassler, K. Sasaki and R.B. Griffiths, J. Stat. Phys.
{\bf 62}, 45 (1992).

\bibitem{Sas92} K. Sasaki, J. Stat. Phys. {\bf 68}, 1013 (1992).


\bibitem{Aubry83a} S. Aubry, Physica {\bf 7D}, 240 (1983).

\bibitem{Aubry83b} S. Aubry and P.Y. Le Daeron, Physica {\bf 8D}, 38 (1983).


\bibitem{Seno94} F. Seno, J.M. Yeomans, R. Harbord and D.K.Y Ko, Phys. Rev.
{\bf B} , to be published (1994).


\bibitem{FS80} M.E. Fisher and W. Selke, Phys. Rev. Lett. {\bf 44},
1502 (1980); Phil. Trans. R. Soc. (London) {\bf 302}, 1 (1981).

\bibitem{Aubry2} S. Aubry, Physics {\bf D71}, 196 (1994).

\bibitem{SRY93} F. Seno, D.A. Rabson and J.M. Yeomans, J. Phys. {\bf A26},
4887 (1993).

\bibitem{Villain} J. Villain and M.B. Gordon, J. Phys. C: Solid State
Physics {\bf 13}, 3117 (1980).

\bibitem{Aubry1} F. Vallet, R. Schilling and S. Aubry, J. Phys. C:
Solid State Physics {\bf 21}, 67 (1988).

\bibitem{SY94} F. Seno and J.M. Yeomans, paper in preparation (1994).

\bibitem{Aubrynew} S. Aubry, J. Phys. C: Solid State Physics {\bf 16},
2497 (1983).

\bibitem{Spzilka} A.M. Spzilka and M.E. Fisher, Phys. Rev. Lett. {\bf 57},
1044 (1986); Phys. Rev. {\bf B36}, 644 (1987).

\end{references}
\end{document}